\documentclass[iop,apjl,tighten]{emulateapj}
\usepackage{graphicx,enumitem}

\shorttitle{A peculiar satellite in the outer halo of M31}
\shortauthors{Mackey et al.}

\begin{document}


\title{A peculiar faint satellite in the remote outer halo of M31\altaffilmark{$\dagger$}}


\author{A. D. Mackey$^1$, A. P. Huxor$^2$, N. F. Martin$^{3,4}$, A. M. N. Ferguson$^5$, A. Dotter$^1$, A. W. McConnachie$^6$,\\R. A. Ibata$^3$, M. J. Irwin$^7$, G. F. Lewis$^8$, C. M. Sakari$^9$, N. R. Tanvir$^{10}$, and K. A. Venn$^9$}
\affil{$^1$\ Research School of Astronomy \& Astrophysics, The Australian National University, Mount Stromlo 
Observatory,\\via Cotter Road, Weston, ACT 2611, Australia; \url{dougal@mso.anu.edu.au}\\
$^2$Astronomisches Rechen-Institut, Universit\"{a}t Heidelberg, M\"{o}nchhofstra{\ss}e 12-14, 69120 Heidelberg, Germany\\
$^3$Observatoire astronomique de Strasbourg, Universit\'{e} de Strasbourg, CNRS, UMR 7550,\\11 rue de l'Universit\'{e}, F-67000 Strasbourg, France\\
$^4$Max-Planck-Institut f\"{u}r Astronomie, K\"{o}nigstuhl 17, 69117 Heidelberg, Germany\\
$^5$Institute for Astronomy, University of Edinburgh, Royal Observatory, Blackford Hill, Edinburgh, EH9 3HJ, UK\\
$^6$NRC Herzberg Institute for Astrophysics, 5071 West Saanich Road, Victoria, BC V9E 2E7, Canada\\
$^7$Institute of Astronomy, University of Cambridge, Madingley Road, Cambridge, CB3 0HA, UK\\
$^8$Sydney Institute for Astronomy, School of Physics, A28, The University of Sydney, NSW 2006, Australia\\
$^9$Department of Physics \& Astronomy, University of Victoria, 3800 Finnerty Road, Victoria, BC V8P 1A1, Canada\\
$^{10}$Department of Physics and Astronomy, University of Leicester, University Road, Leicester, LE1 7RH, UK}

\altaffiltext{$\dagger$}{Based on observations made with the NASA/ESA {\it Hubble Space Telescope}, obtained at the Space Telescope Science Institute (STScI), which is operated by the Association of Universities for Research in Astronomy, Inc., under NASA contract NAS 5-26555. These observations are associated with program GO 12515.}


\begin{abstract}
We present {\it Hubble Space Telescope} imaging of a newly-discovered faint stellar system, PAndAS-48, in the outskirts of the 
M31 halo. Our photometry reveals this object to be comprised of an ancient and very metal-poor stellar population with 
age$\,\ga10$\ Gyr and $[$Fe$/$H$]\la-2.3$. Our inferred distance modulus $(m-M)_0=24.57\pm0.11$ confirms that PAndAS-48 
is most likely a remote M31 satellite with a 3D galactocentric radius of $149^{+19}_{-8}$\ kpc. We observe an apparent spread in color 
on the upper red giant branch that is larger than the photometric uncertainties should allow, and briefly explore the implications of this.
Structurally, PAndAS-48 is diffuse, faint, and moderately flattened, with a half-light radius $r_h=26^{+4}_{-3}$\ pc, integrated 
luminosity $M_V=-4.8\pm0.5$, and ellipticity $\epsilon=0.30^{+0.08}_{-0.15}$. On the size-luminosity plane it falls between the 
extended globular clusters seen in several nearby galaxies, and the recently-discovered faint dwarf satellites of the Milky Way; 
however, its characteristics do not allow us to unambiguously class it as either type of system. If PAndAS-48 is a globular cluster 
then it is the among the most elliptical, isolated, and metal-poor of any seen in the Local Group, extended or otherwise. Conversely, 
while its properties are generally consistent with those observed for the faint Milky Way dwarfs, it would be a factor $\sim2-3$ 
smaller in spatial extent than any known counterpart of comparable luminosity.
\end{abstract}




\section{Introduction}
In recent years the advent of deep wide-field sky surveys has heralded the discovery of a variety of new
and unusual diffuse stellar systems in the Local Group. Arguably the most significant of these are the 
ultra-faint dwarf (UFD) satellites of the Milky Way \citep[e.g.,][]{willman:05,zucker:06,belokurov:07}, 
the existence of which may have important implications for the ``missing satellites'' problem of $\Lambda$CDM 
cosmology \citep[e.g.,][]{koposov:09}. These objects have luminosities as low as $M_V\sim-1.5$, 
and, with characteristic radii commonly in the range $r_h\sim25-100$\ pc \citep[e.g.,][]{martin:08,sand:12}, 
physical sizes that are substantially smaller than those of classical dwarf spheroidal galaxies and which in 
some cases approach the globular cluster (GC) regime.

UFDs have two key characteristics distinguishing them as galaxies. First, although analysis of their internal
kinematics is fraught with complexity due to, for example, the presence of binary stars and the effects of external
tidal forces \citep[e.g.,][]{mcconnachie:10,simon:11}, radial velocity observations 
consistently imply mass-to-light ratios as high as a few thousand \citep[e.g.,][]{martin:07,simon:07}.
Second, their constituent stellar populations (i) are ancient, and very metal poor with 
$\langle [$Fe$/$H$] \rangle\la-2.2$ \citep[e.g.,][]{kirby:11,brown:12}; (ii) frequently exhibit substantial 
internal dispersions in iron abundance of up to $\sim0.6-0.7$ dex \citep[e.g.,][]{martin:07,kirby:08}; and 
(iii) sometimes include extremely metal poor members with $[$Fe$/$H$]<-3$ \citep[e.g.,][]{norris:10}. 
UFDs apparently extend the metallicity-luminosity relation seen for classical Milky Way satellites 
\citep[e.g.,][]{kirby:11}, suggesting that they are not the stripped remnants of once-larger progenitors
\citep[see also][]{leaman:12}. 

\begin{figure*}
\begin{center}
\includegraphics[width=140mm]{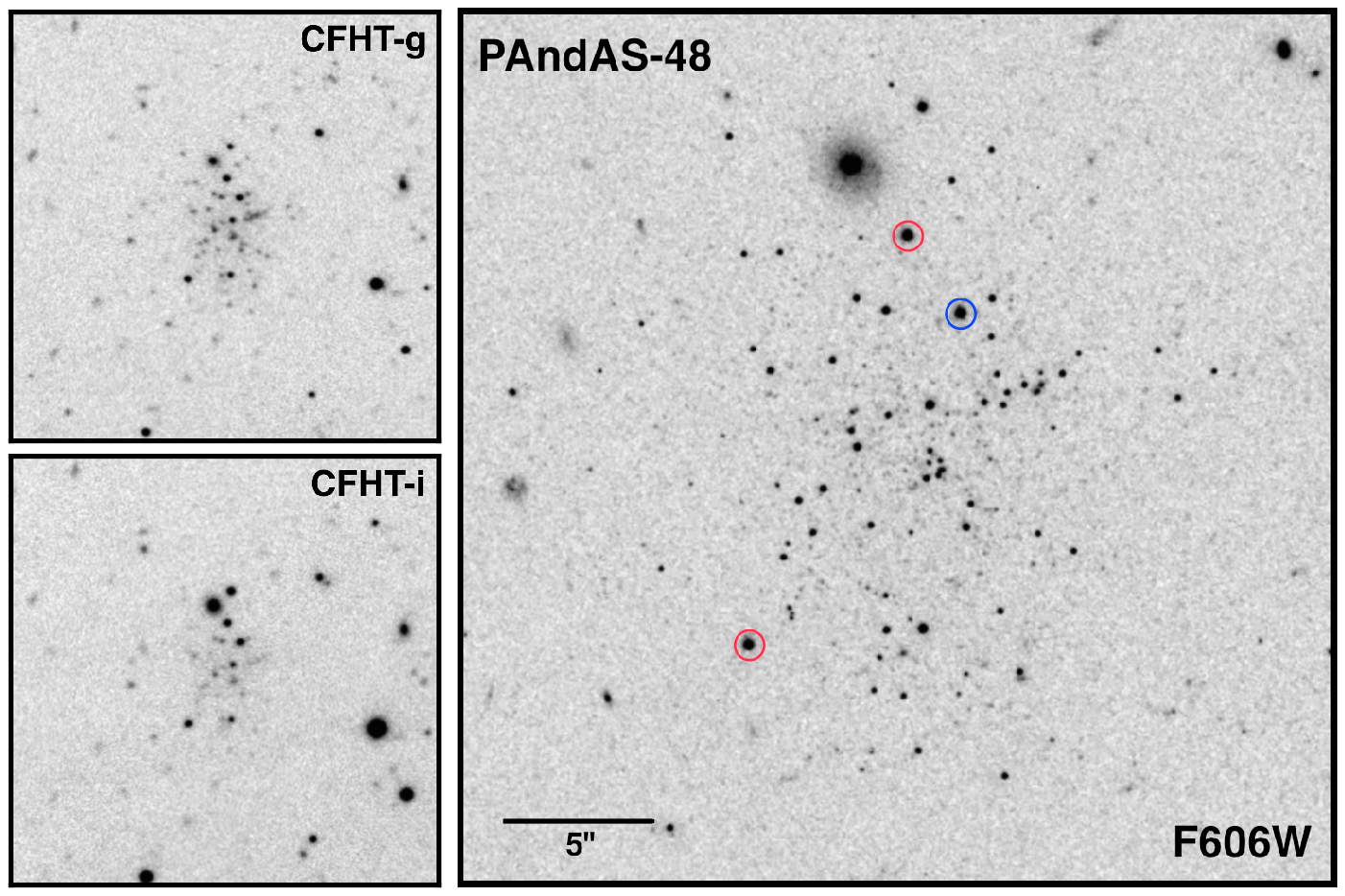}
\end{center}
\caption{{\bf Left panels:} PA-48 discovery images, $1\arcmin$ on a side, in the CFHT/MegaCam $g$- and $i$-bands. {\bf Right panel:} Our $30\arcsec\times30\arcsec$ drizzled F606W ACS/WFC image of PA-48. Star-A and star-C from Section \ref{ss:cmd} are circled in red; star-B in blue. North is to the top and east to the left in all images.\label{f:images}}
\end{figure*}
 
A second recently-discovered class of diffuse stellar systems is that of the so-called ``extended clusters''
\cite[ECs;][]{huxor:05}. These provoked interest because the first luminous examples 
appeared isolated on the size-luminosity plane, encroaching upon the empty region separating GCs from 
dwarf galaxies. However subsequent work uncovered many fainter ECs in M31 \citep{huxor:08} and other
Local Group galaxies \citep{stonkute:08,huxor:09,huxor:13,hwang:11},
and these fill out a region on the plane with $-8\la M_V\la-5$ and $r_h\sim15-35$\ pc that overlaps 
substantially with the most diffuse GCs seen in the Milky Way \citep{huxor:11}. Along with refined structural 
measurements \citep{tanvir:12} and resolved photometry precluding large internal spreads in $[$Fe$/$H$]$ 
\citep{mackey:06}, this suggests that ECs most likely represent the upper tail of the globular cluster size 
distribution \citep[see also][]{dacosta:09}.
It remains to be convincingly demonstrated kinematically 
that ECs need not contain any dark matter component \citep[e.g.,][]{collins:09}, but they nonetheless apparently 
stand clearly distinct from UFDs despite their close proximity on the size-luminosity plane.

M31 is an excellent location for studying a wide variety of stellar systems, as it possesses many 
more dwarf satellites \citep[e.g.,][]{mcconnachie:12} and globular clusters \citep[e.g.,][]{galleti:04} 
than does the Milky Way.
Recently we have conducted the {\it Pan-Andromeda Archaeological Survey} \citep[PAndAS;][]{mcconnachie:09}, 
imaging the M31 halo to a projected galactocentric distance of $R_p\approx150$\ kpc with CFHT/MegaCam and
facilitating the first detailed characterization of this region and the star clusters and dwarf galaxies that inhabit it 
\citep[e.g.,][]{huxor:08,mcconnachie:08,richardson:11}.

\begin{figure*}
\begin{center}
\includegraphics[width=155mm]{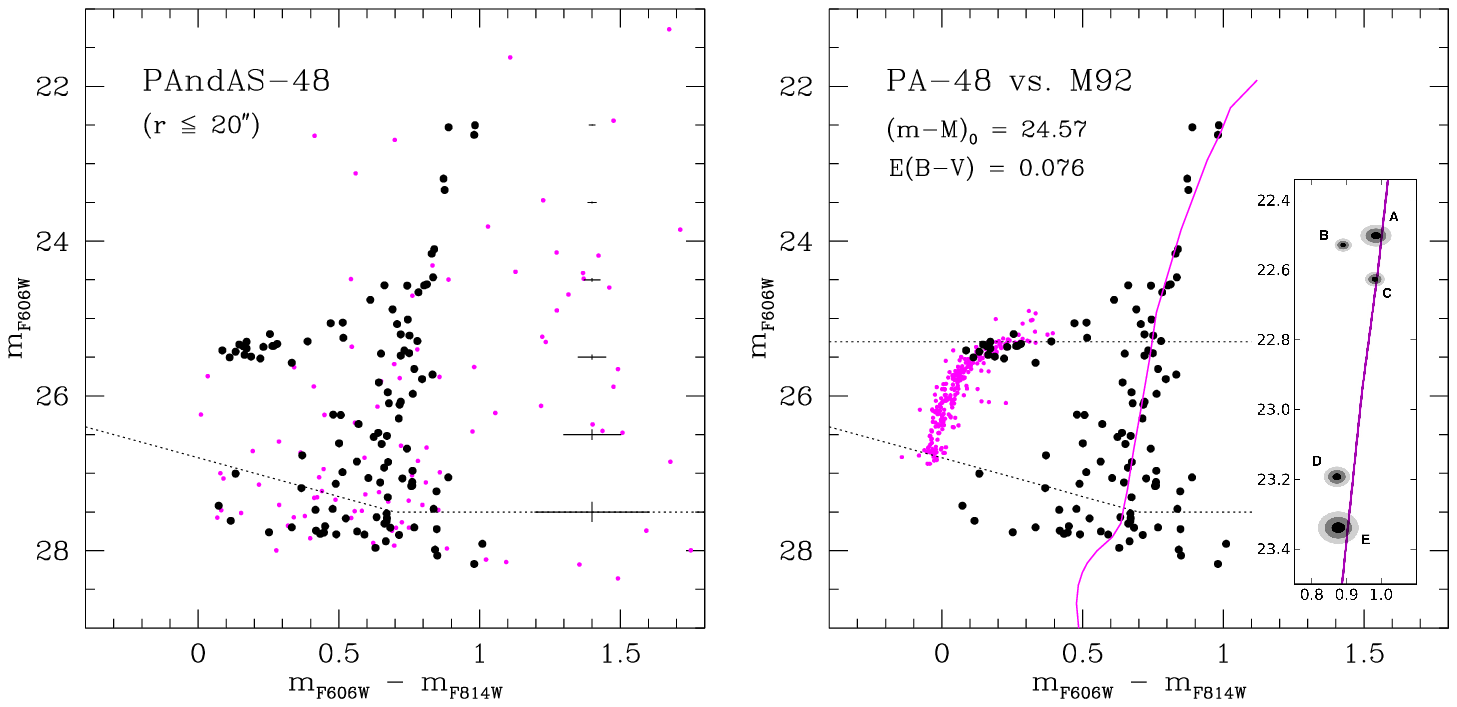}
\end{center}
\caption{{\bf Left panel:} Our ACS/WFC CMD for PA-48. Field stars more than $50\arcsec$ away from the cluster center are marked with small magenta points. The dotted line is our $50\%$ completeness level. {\bf Right panel:} The PA-48 CMD with registered M92 fiducial and HB. The inset shows stars on the upper RGB with individual $1\sigma$, $2\sigma$, and $3\sigma$ photometric uncertainties. The five brightest stars are labelled A-E.\label{f:cmds}}
\end{figure*}
 
One of the most intriguing of our discoveries is PAndAS-48 (PA-48), at $\alpha=00^{\rm h}59^{\rm m}28\fs3$, 
$\delta=+31\degr29\arcmin10\farcs6$ (J2000), which we uncovered during our search for remote globular clusters in M31
(Huxor et al.\ 2013, in prep.). This is a very faint stellar system with $M_V\approx-4.7$ lying at $R_p=141$\ kpc. 
The PAndAS discovery images (Figure~\ref{f:images}) reveal a notably elongated object with a very diffuse structure.
In this Letter we present deep {\it Hubble Space Telescope (HST)} follow-up imaging of 
PA-48 and show that, unusually, it possesses characteristics that do not allow it to be unambiguously 
classified as either an extended cluster or a faint dwarf galaxy.

\section{Observations and data analysis}
We observed PA-48 with the Wide Field Channel (WFC) of the Advanced Camera for Surveys
(ACS) aboard {\it HST} on 2011 November 3 as part of program GO 12515 (PI: Mackey). The object was
imaged three times in both the F606W and F814W filters, with small dithers between exposures.
Integration times were $799$s for F606W frames and $845$s for F814W frames. 

We obtained the reduced data from the {\it HST} archive. The {\sc calacs} pipeline now includes a pixel-based 
charge-transfer efficiency (CTE) correction \citep[e.g.,][]{anderson:10} as well as improved corrections for effects 
introduced by the ACS electronics. In the right panel of Figure~\ref{f:images} we display the drizzled, CTE-corrected 
F606W image of PA-48. The object is fully resolved, and we confirm it as a faint, low surface brightness stellar 
system with an extended structure; the apparently elongated morphology persists.

We photometered our images using v2.0 of the {\sc dolphot} photometry software \citep{dolphin:00}.
{\sc Dolphot} performs point-spread function (PSF) fitting using model PSFs especially tailored to the ACS/WFC camera. 
The software provides a variety of photometric quality parameters 
for each detection. We selected those objects classified as stellar, with valid photometry in all six input images, 
a global sharpness parameter between $\pm0.1$ in each filter, and a crowding parameter $\leq0.08$ mag in each filter. 
Our final photometry is on the calibrated VEGAMAG scale of \citet{sirianni:05}. 

We assessed the photometric uncertainties and detection completeness by performing artificial star tests.
For each real star we randomly generated $500$ artificial stars of the same brightness but with 
positions uniformly distributed within a radius $1\arcsec\leq r\leq5\arcsec$ of its coordinates. {\sc dolphot}
adds a single artificial star at a time to the images and then attempts to measure it. We filtered the results according 
to the photometric quality parameters as described above, and for each real star (i) assigned a detection completeness 
by comparing the number of recovered objects with the number submitted; and (ii) characterised the photometric 
uncertainties by examining the scatter in the differences between the input and measured magnitudes of the artificial 
stars. Our $50\%$ completeness level is at $m_{{\rm F606W}}=27.5$ and $m_{{\rm F814W}}=26.8$, where the uncertainties 
are typically $\pm0.1$ mag in both filters. 

\section{Results}
\subsection{Color-magnitude diagram}
\label{ss:cmd}
Figure~\ref{f:cmds} shows our color-magnitude diagram (CMD) for PA-48. We observe a steep red giant branch
(RGB) and a blue horizontal branch (HB), characteristic of the ancient ($\ga10$\ Gyr old) stellar populations 
seen in both GCs and low-luminosity dwarfs. At the top of the RGB there is the suggestion of a spread
in color that is larger than the photometric uncertainties should allow; we discuss this in more detail below. 

To obtain a photometric metallicity estimate for PA-48 we aligned de-reddened globular cluster fiducials 
from \citet{brown:05}, following the procedures described in \citet{mackey:06,mackey:07}.
Briefly, we registered the fiducials using the observed level of the HB and the color of the
RGB at the HB level. The metallicity sets the curvature of the upper RGB. Because PA-48 is in the remote 
halo of M31, it may not lie at the same distance as the galaxy's center. For each fiducial we therefore tested 
distance moduli in a range $\pm0.5$ mag about the usual $(m-M)_0=24.47$ for M31. At given $(m-M)_0$
we calculated the foreground extinction necessary to align the HB levels of the CMD and the 
reference fiducial. This in turn allowed us to calculate the offset in color between the CMD and fiducial 
RGB sequences at the HB level. The best combination of $(m-M)_0$ and $E(B-V)$ for a given fiducial was 
that which minimised this offset. We then examined how closely the shapes of the CMD and fiducial RGB 
sequences matched.
 
\begin{figure*}
\begin{center}
\includegraphics[width=155mm]{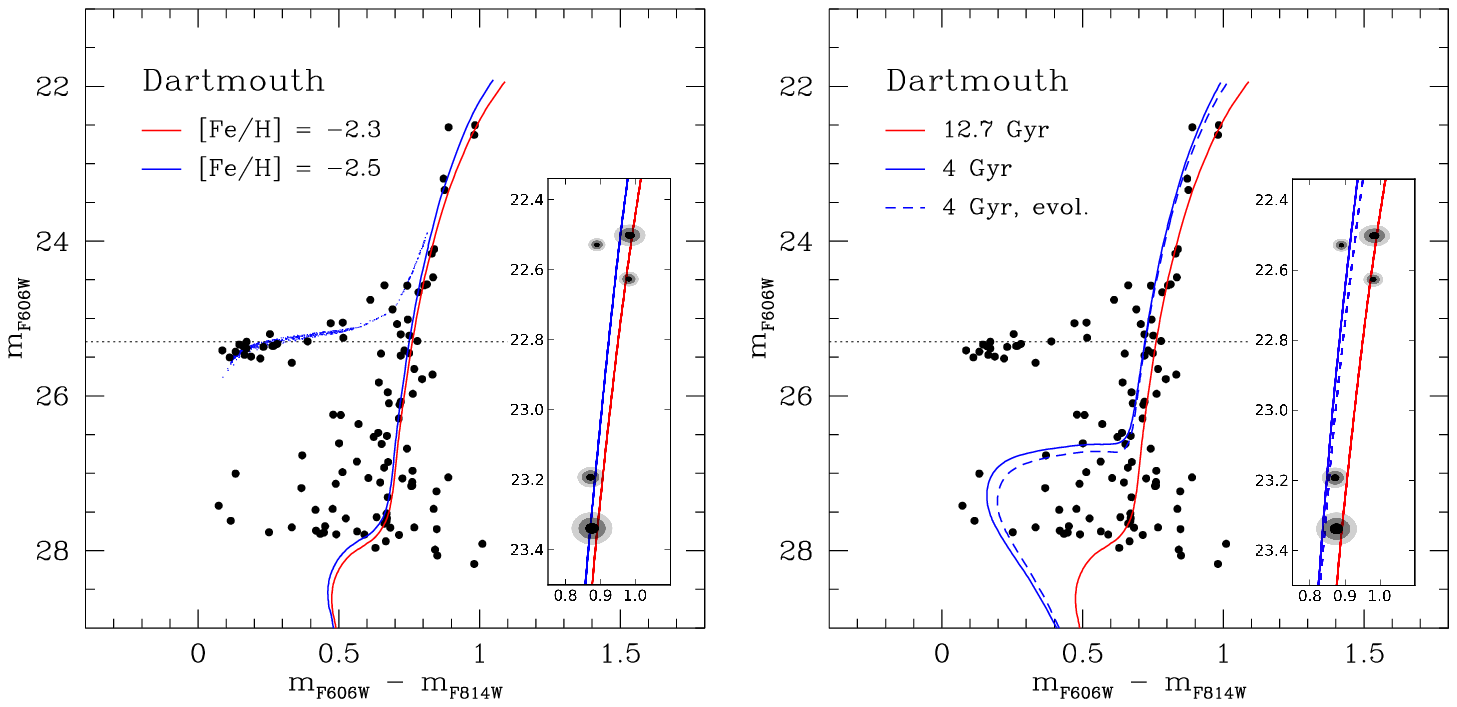}
\end{center}
\caption{{\bf Left panel:} The PA-48 CMD with registered Dartmouth isochrones as labelled. Both have $[\alpha/$Fe$]=+0.4$. For clarity only the HB for the $[$Fe$/$H$]=-2.5$ model is plotted. {\bf Right panel:} Registered isochrones showing the effects of an internal age dispersion. The first two models have $[$Fe$/$H$]=-2.3$ and $[\alpha/$Fe$]=+0.4$; the third ``chemically evolved'' model has $[$Fe$/$H$]=-1.9$ and $[\alpha/$Fe$]=+0.2$\label{f:iso}}
\end{figure*}

The most metal-poor cluster in the \citet{brown:05} sample -- M92, with $[$Fe$/$H$]\approx-2.3$
\citep[see][2012 update]{harris:96} -- provides an excellent fit to the shape of the red side 
of the PA-48 RGB. This suggests that PA-48 is at least as metal-poor as M92. The implied distance modulus 
and foreground reddening are $(m-M)_0=24.57\pm0.11$ and $E(B-V)=0.08\pm0.01$, where the
uncertainties are calculated according to the prescription of \citet{mackey:06,mackey:07}. 
Our reddening estimate is in good agreement with that predicted by the \citet{schlegel:98} maps, $E(B-V)=0.066$,
while our distance modulus strongly suggests PA-48 is a member of the M31 system -- it sits at $820^{+43}_{-41}$ kpc, 
implying a 3D galactocentric radius of $149^{+19}_{-8}$\ kpc. The PA-48 HB does not extend bluewards as far as that
of M92. We measure a HB-index $(B-R)/(B+V+R)\approx0.6$, which may indicate a mild second-parameter effect.
%

The inset to Figure~\ref{f:cmds} highlights the fact that stars on the upper RGB of PA-48 span a range in 
color larger than would be expected if considering only the photometric uncertainties. The registered M92 fiducial
suggests that this is driven predominantly by the bluest, and second brightest, of the three most luminous RGB stars 
(hereafter star-B). Given the potential importance of an RGB color spread, we consider 
several possibilities that might explain the observed position of star-B on the CMD:

{\it 1. Non-member.} Local foreground contamination is moderate -- we find $22$ stars further than $20\arcsec$ from
the PA-48 center that are at least as bright as its upper RGB ($m_{\rm F606W}\la23.5$; Figure~\ref{f:cmds}). However, star-B 
lies well within the half-light radius of $r_{h}\approx6.6\arcsec$ (see below). The probability that a non-member should 
fall both within $r_h$ and close to the RGB is very small. The area within $r_h$ is $0.34\%$ of the WFC field-of-view, while 
the vicinity of the RGB is, generously, $\approx\slantfrac{1}{12}$ of the area of the CMD occupied by stars with $m_{\rm F606W}\la23.5$. 
Assuming field stars are uniformly distributed on the CCD and the CMD, the chance of one or more falling both within 
$r_h$ and close to the RGB from $23$ trials is just $0.68\%$. Hence it is very likely that star-B is a member of PA-48.


{\it 2. An internal dispersion in $[$Fe$/$H$]$ or age.} In principle, an internal dispersion in $[$Fe$/$H$]$ ought to 
translate into a color spread on the upper RGB, with star-B's blue $m_{\rm F606W}-m_{\rm F814W}$ magnitude 
indicating a lower abundance than for star-A and star-C. 
However, stellar evolution models generally indicate that at fixed luminosity 
the RGB temperature and, hence, broadband color will asymptote to a certain value with decreasing metallicity.
To test the sensitivity of $m_{\rm F606W}-m_{\rm F814W}$ color at low $[$Fe$/$H$]$ we use isochrones from the
Dartmouth Stellar Evolution Database \citep{dotter:08}. In Figure~\ref{f:iso} we plot models with 
$[$Fe$/$H$]=-2.3$\ and $-2.5$, and $[\alpha/$Fe$]=+0.4$. There is a noticeable shift to the blue with
decreasing abundance, but not sufficiently large to match star-B's color. We further computed Dartmouth
evolutionary tracks at $0.8M_{\odot}$ and $[\alpha/$Fe$]=+0.4$ for $[$Fe$/$H$]=-2.5$, $-3.0$, and $-3.5$. 
On the upper RGB at a luminosity consistent with that of star-B these three models differ in F606W$-$F814W 
color by no more than $0.01$ mag. Thus, under the assumption of constant $[\alpha/$Fe$]$, it is unlikely that
star-B's position on the CMD is due to it having a much lower $[$Fe$/$H$]$ than the bulk of the other stars.
However, the insensitivity of $m_{\rm F606W}-m_{\rm F814W}$ color to $[$Fe$/$H$]$ at very low abundances 
also means that we cannot rule out a moderate internal metallicity dispersion in general.

A complicating factor is that our photometry does not reach the $\sim 13$ Gyr main-sequence turn-off; we can 
only exclude the presence of populations younger than $\approx 4$ Gyr (Figure~\ref{f:iso}).
Decreasing age at fixed composition moves the RGB to the blue; $4$ Gyr old stars might nearly match the 
position of star-B. Mild chemical evolution with age could also be accommodated -- as an example
we plot a $4$ Gyr model with $[$Fe$/$H$]=-1.9$ and $[\alpha/$Fe$]=+0.2$. 

{\it 3. Asymptotic giant branch (AGB) star.} The time a low-mass star spends ascending the AGB (i.e., before 
the thermally-pulsating phase) is $\sim10\%$ of the time it spends in the core He-burning phase. 
Thus we may expect to find roughly one AGB star for every ten HB stars in the CMD. In PA-48 we observe
$\sim17-20$ HB stars, consistent with the possibility that star-B is an AGB star. Even so, this object is 
substantially bluer than all AGB stars visible at comparable brightness in the well-populated CMDs for metal-poor 
M31 ECs presented by \citet{mackey:06}, which tend to lie much closer to the RGB. It is hence difficult to 
conclusively assess the viability of this suggestion with only the presently-available data; however, in our
opinion it remains the most likely option.

\subsection{Structure \& luminosity}
\begin{figure}
\begin{center}
\includegraphics[width=80mm]{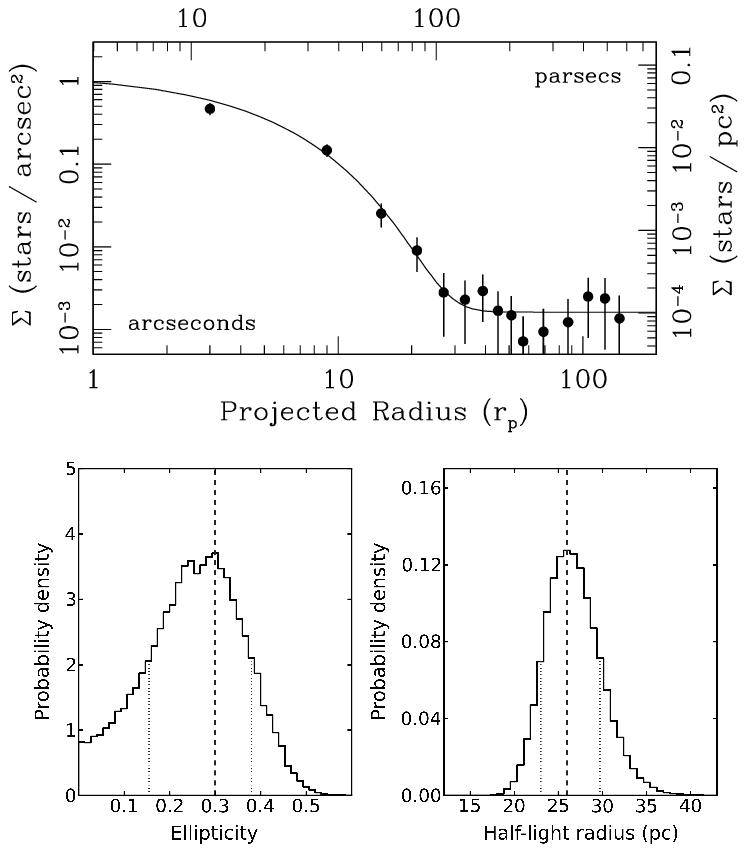}
\end{center}
\caption{{\bf Upper panel:} Radial surface density profile for PA-48, where the points are the stellar density measured in elliptical annuli and the line represents the best-fitting model found using the algorithm of \citet{martin:08}. {\bf Lower panels:} Marginalized posterior probability distribution functions for the ellipticity and half-light radius of PA-48. The most likely values are indicated with a dashed line; $1\sigma$ uncertainties are marked with dotted lines.\label{f:profile}}
\end{figure}

To quantify the structure of PA-48 we utilised the maximum likelihood algorithm developed by
\citet{martin:08} to study low-luminosity stellar systems. This technique provides robust estimates and uncertainties 
for parameters such as the half-light radius and ellipticity of an object from resolved photometry.
Our results are displayed in Figure~\ref{f:profile}. The algorithm converges on a unique likelihood maximum, 
revealing that PA-48 is indeed a moderately elliptical extended object with $\epsilon=0.30^{+0.08}_{-0.15}$ 
and $r_h=6.6^{+0.9}_{-0.8}\arcsec\approx26^{+4}_{-3}$\ pc. The uncertainties represent $1\sigma$ confidence
limits; note there is an additional $\pm 1$\ pc uncertaintiy in $r_h$ due to the error in our line-of-sight distance.

Huxor et al.\ (2013, in prep.) estimate $M_V\approx-4.7$ for PA-48 from surface photometry of the discovery images.
With our resolved measurements we are able to integrate to a larger radius and use a CMD filter to better 
remove contaminants. The completeness-corrected luminosity we obtain by adding all
members down to $m_{\rm F606W}=27.5$ and within $30\arcsec$ of the center is $M_V=-4.3$. Experimentation with population
synthesis models suggests that for a \citet{kroupa:01} mass function $\approx65\%$ of the total luminosity of
the system is included to our faint limit, implying $M_V\approx-4.8$. As per \citet{martin:08}, uncertainties in the
total luminosity for systems this faint are dominated by CMD shot noise at a level $\pm0.5$\ mag.

\section{Discussion}
\begin{figure*}
\begin{center}
\includegraphics[width=128mm]{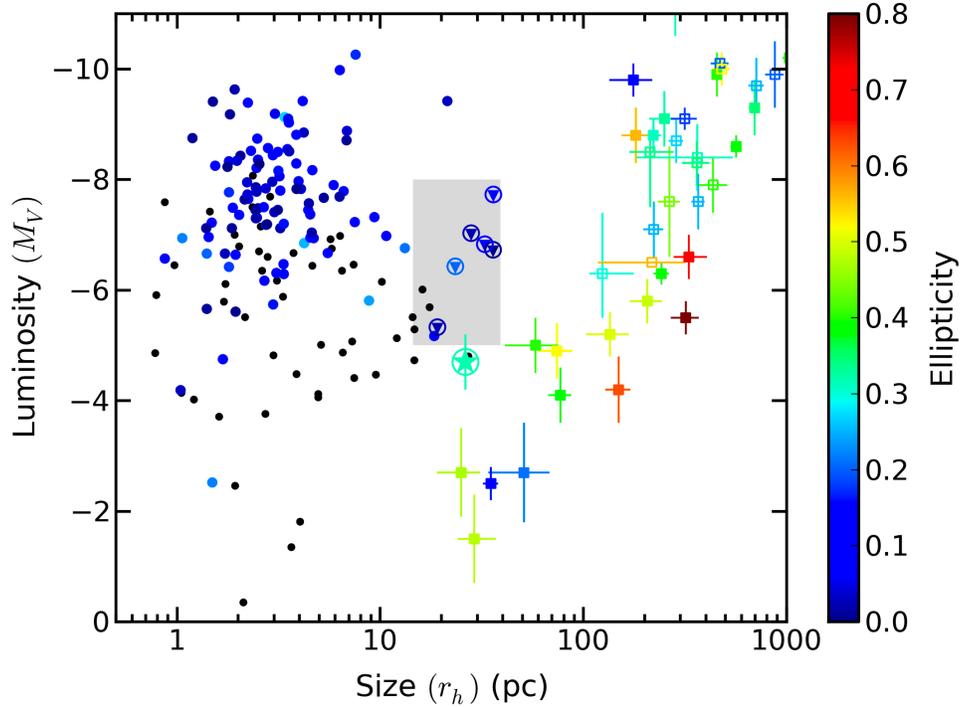}
\end{center}
\caption{Luminosity versus size for a variety of stellar systems in the Local Group, color-coded by ellipticity. Filled circles are Galactic globular clusters \citep{harris:96}; those with no ellipticity measurement are marked in black. Filled (empty) squares are Galactic (M31) dwarf satellites \citep{mcconnachie:12}, while circled triangles are M31 ECs (this work). The grey shaded region denotes the approximate area occupied by ECs observed in various Local Group galaxies \citep[e.g.,][]{huxor:11}. PA-48 is marked with a circled star.\label{f:sizelum}}
\end{figure*}

PA-48 is an unusual M31 satellite that is not easily classified. 
Structurally it is similar to the faint dwarf companions of the Milky Way as well as the ECs seen around some 
galaxies in the Local Group. Intriguingly, however, a number of its other characteristics would make it an atypical 
example of either type of object. To illustrate this point, in Figure~\ref{f:sizelum} we plot a variety of stellar 
systems on the size-luminosity plane, color-coded by ellipticity. Assessed purely in terms of $r_h$ and $M_V$, 
PA-48 appears most akin to the lowest-luminosity Local Group ECs. Although the faintest Milky Way 
UFDs are of very similar size to PA-48, dwarfs of comparable luminosity to PA-48 have scale radii typically 
$\sim 2-3$ times larger. 

That said, PA-48 appears substantially more elliptical than any Galactic globular cluster measured to date, extended or
otherwise, as well as almost all Local Group GCs for which reliable measurements are available \citep[Figure~6 in][]{huxor:13}.
However, it lies comfortably within the range of ellipticities exhibited by faint dwarfs \citep[e.g.,][]{martin:08,sand:12}. 
Unfortunately, few of the low-luminosity diffuse Milky Way globular clusters have ellipticity measurements in the 
\citet{harris:96} catalogue. To attempt a fairer comparison, we ran the \citet{martin:08} software on our 
{\it HST}\ photometry of M31 ECs -- three from 
\citet{mackey:06}\footnote{We excluded EC3$\,\equiv\,$HEC4 due to heavy field contamination.},
and three from the present program, yet unpublished\footnote{PA-2, PA-12, and PA-50.}.
These span luminosities $-7.7\la M_V\la-5.3$ and projected galactocentric radii $15\la R_{\rm p}\la115$\ kpc.
Our results are consistent with the assessment above. Five of the six M31 ECs have $\epsilon<0.1$ with typical
uncertainties of $\pm0.08$, while one (PA-50) has $\epsilon=0.20\pm0.07$, commensurate with the most highly 
flattened Galactic GCs. 

If PA-48 is a cluster then it would also be among the most isolated such objects seen in the Local Group -- just one GC 
of any kind (MGC1 in M31) is known with a larger 3D radius. This appears at odds with its unusually high ellipticity.
One might expect that a low-luminosity star cluster with high ellipticity is being tidally disturbed; however we have strong
evidence from MGC1 that tidal forces are exceedingly benign at such large galactocentric distances \citep{mackey:10}. 
It may be that PA-48 is on a highly eccentric orbit about M31 and so suffered a disruptive but non-fatal tidal shock some 
Gyr ago. A radial velocity measurement and orbital modelling will be necessary to determine if this is a viable solution. 

As noted above, if PA-48 is a low-luminosity dwarf then it is unusually small in terms of its spatial extent.
Comparable systems are observed (Willman 1; Segue 1 \& 2) but these are $\sim2-3$\ mag fainter than PA-48.
Although the luminosities of such objects can fluctuate substantially with the evolution of their brightest few stars, it is
very unlikely they could get as bright as PA-48 \citep[or vice versa, see][]{martin:08}; however, the distinction between these
systems may not be as great as it appears on Figure~\ref{f:sizelum}. Furthermore, only a handful of dwarf 
galaxies fainter than $M_V=-5$ are known and it is easy to speculate that perhaps the size-luminosity plane is simply not 
yet sufficiently well sampled to accurately assess how unusual PA-48 would be in this regard. For dwarf galaxies with 
$M_V\la-8$, where completeness is less of an issue, there is nearly an order of magnitude spread in $r_h$ at given 
luminosity \citep[e.g.,][]{mcconnachie:12}.

A key alternative characteristic defining faint dwarfs is an internal spread in 
$[$Fe$/$H$]$. Unfortunately, we are not able to constrain this possibility for PA-48 
given its very metal-poor nature and our choice of broadband filters. \citet{kirby:11} have shown that UFDs follow 
the same metallicity-luminosity relation as do the more massive Galactic satellites (their Figure~3). Intriguingly, with 
$\langle[$Fe$/$H$]\rangle\la-2.3$ and $\log(L/L_\odot)\approx3.8$, PA-48 fits closely with this relationship.
If PA-48 is a dwarf galaxy then it would be the faintest known example outside the Milky Way. Its isolation 
would reinforce the possibility that the faintest Galactic UFDs could well exist out to comparable distances, 
and not be shaped through their interactions with their host.

Finally, it is relevant that PA-48 lies within the thin plane of corotating M31 dwarf galaxies recently
identified by \citet{ibata:13} \citep[see also][]{conn:13}. In this context, knowledge of its radial velocity would be
helpful -- if PA-48 is a true member of this plane of satellites (PoS) then we expect its velocity to be more negative 
than the M31 systemic velocity. One might suspect that membership of the M31 PoS would cement PA-48's status 
as a dwarf galaxy, but this is not clear cut. We do not observe any evidence for M31 GCs being associated with
the plane; however, \citet{keller:12} argue that outer halo Milky Way GCs {\it are} members of 
the Galactic PoS defined by, e.g., \citet{metz:09}.

\acknowledgements
ADM and GFL are grateful for support from the Australian Research Council (Discovery Projects DP1093431 
\& DP110100678, and Future Fellowship FT100100268).

{\it Facilities:} \facility{HST}.

\end{document}